\documentclass[conference]{IEEEtran}
\ifCLASSINFOpdf
\else
\fi

\usepackage{comment}
\usepackage{graphicx}
\usepackage{caption}
\usepackage{subcaption}
\usepackage{makecell}
\usepackage{amsmath,amssymb,amsfonts}
\usepackage{todonotes}

\IEEEoverridecommandlockouts
\usepackage{tikz}
\usepackage{textcomp}
\usepackage[hidelinks]{hyperref}
\usepackage{lipsum}
\newcommand\copyrighttext{%
  \footnotesize \textcopyright 2026 IEEE. Personal use of this material is permitted.  Permission from IEEE must be obtained for all other uses, in any current or future  media, including reprinting/republishing this material for advertising or promotional  purposes, creating new collective works, for resale or redistribution to servers or  lists, or reuse of any copyrighted component of this work in other works.
  DOI: TBD %\href{<http://tex.stackexchange.com>}{TBD}
  }
\newcommand\copyrightnotice{%
\begin{tikzpicture}[remember picture,overlay]
\node[anchor=south,yshift=10pt] at (current page.south) {\fbox{\parbox{\dimexpr\textwidth-\fboxsep-\fboxrule\relax}{\copyrighttext}}};
\end{tikzpicture}%
}
\begin{document}
\title{Impact of Optical Reconfigurable Intelligent Surfaces on Channel Estimation for VLC Systems}

%\author{\IEEEauthorblockN{Borja Genoves Guzman}
%\IEEEauthorblockA{Departamento de Teoría de la Señal y Comunicaciones\\
%Leganés, Madrid, Spain 28911\\
%Email: bgenoves@ing.uc3m.es}
%\and
%\IEEEauthorblockN{M. Julia Fernández-Getino García}
%\IEEEauthorblockA{Departamento de Teoría de la Señal y Comunicaciones\\
%Leganés, Madrid, Spain 28911\\
%Email: mjfgarci@ing.uc3m.es}
%}

\author{
\IEEEauthorblockN{Borja Genoves Guzman and M. Julia Fernández-Getino García}
\IEEEauthorblockA{Signal Theory and Communications Department, Universidad Carlos III de Madrid, Leganés, Madrid 28911 Spain}
\IEEEauthorblockA{E-mails: bgenoves@ing.uc3m.es, mjfgarci@ing.uc3m.es}
}


%Telephone: (800) 555--1212, Fax: (888) 555--1212}


\maketitle

\copyrightnotice 
\vspace{-\baselineskip}

\begin{abstract}
Visible light communication (VLC) offers wireless data connectivity indoors, yet its performance is severely limited by link blockages. While optical reconfigurable intelligent surfaces (ORISs) have been proposed to overcome these blockages, existing literature predominantly relies on the unrealistic assumption of ideal channel state information. In this paper we investigate the contribution of ORIS elements in VLC systems under practical, non-ideal channel estimation scenarios. We analyze the communication performance using two pilot-based estimation techniques: pilot symbol assisted modulation (PSAM) and superimposed training (ST), evaluating the impact of data power allocation factors and averaged orthogonal frequency division multiplexing (OFDM) symbols. %We demonstrate that deploying ORIS elements yields a considerable signal-to-noise ratio (SNR) improvement, effectively mitigating line-of-sight blockages. 
Our results show that PSAM with an ORIS deployment enables optimal  large data power allocation factors, resulting in spectral efficiency improvements that surpass scenarios with ideal channel estimates but without ORIS deployment. While ST performs well with a large number of averaged OFDM symbols, it is unable to overcome the performance of PSAM when ORIS elements are deployed. We also find the system robust to imperfect ORIS orientation; even with only half of the elements accurately aligned, performance remains acceptable. Ultimately, we show that spectral efficiency values of approximately 5 bits/s/Hz are achievable using ORIS with PSAM, a performance comparable to, or even exceeding, that of unobstructed line-of-sight links.\end{abstract}


% \fi
\IEEEpeerreviewmaketitle

\section{Introduction}
\label{sec:Intro}
Recent advances in visible light communication (VLC) have positioned it as a high-bandwidth, secure, and cost-effective alternative to traditional radio frequency (RF) systems, particularly for indoor environments~\cite{VLC}. By leveraging the existing lighting infrastructure, VLC systems utilize light emitting diodes (LEDs) to simultaneously provide illumination and high-speed data transmission through techniques such as direct current biased optical orthogonal frequency division multiplexing (DCO-OFDM)~\cite{RateAndIlluminationVLC}. This intrinsically creates a multiple-input-single-output (MISO) VLC system where multiple LEDs can simultaneously cooperate to serve a user located in a room. This system strongly depends on the presence of line-of-sight (LoS) between transmitter and receiver, which is difficult to guarantee in dynamic scenarios where users, other people, or objects are moving around.

To address LoS limitations, optical reconfigurable intelligent surfaces (ORIS) have emerged as a transformative technology~\cite{ORIS_VLC}. Unlike their RF counterparts, ORIS elements, often implemented via metasurfaces or mirrors~\cite{MirrorsVSMEtasurfaces}, allow for the spatial control of the optical beam, effectively reconfiguring the propagation environment to enhance signal reception. While metasurfaces offer significant promise as wall-mounted ORIS components to boost receiver efficiency in VLC systems~\cite{Metasurfaces1, Metasurfaces2}, their commercial availability remains quite limited. In contrast, the superior reflective properties of mirrors, combined with recent advancements in micro-electromechanical systems (MEMS) for precise orientation control, establish mirrors as the most practical and immediate material for ORIS applications in VLC today \cite{MirrorsVSMEtasurfaces}. Prior works focused on exploiting the advantages of ORIS deployment in signal-to-noise-power ratio (SNR) improvement for an outage probability minimization~\cite{Mirrors1, Mirrors2, Mirrors3}, secrecy rate~\cite{ORISsecrecy_rate} or spectral efficiency maximization~\cite{ORIS_VLC}. However, all prior works assume a perfect channel state information knowledge. This work analyzes the influence of ORIS deployment in a realistic channel estimation dependent MISO-VLC scenario.

There are two main pilot-based channel estimation techniques in optical OFDM: pilot symbol assisted modulation (PSAM) and superimposed training (ST). In PSAM-based systems, dedicated subcarriers or time slots are exclusively reserved for training symbols, which, while providing high accuracy, results in a proportional loss of spectral efficiency~\cite{PSAM_VLC, PSAM_VLC2}. Conversely, ST techniques overlay the training sequence onto the data symbols, preserving the transmission bandwidth at the cost of potential interference between the data and the pilot sequence~\cite{ST_VLC1, ST_VLC2, ST_VLC3}. While prior works have explored the superiority of ST in specific MISO and SISO-VLC contexts, the integration of ORIS elements introduces unique channel characteristics that may shift the comparative advantage of these techniques.

This paper provides a comprehensive evaluation of both PSAM and ST channel estimation techniques within an ORIS-aided MISO-VLC environment. We analyze how array size and orientation inaccuracy in ORIS elements impact SNR and spectral efficiency. By characterizing the trade-offs between these two estimation paradigms, this study aims to provide a framework for selecting the optimal estimation strategy based on the ORIS array size and the required quality of service.

%The structure of the paper is as follows. Section\,\ref{sec:SystemModel} introduces the system model, including LoS and non-LoS (NLoS) channel gains. Section\,\ref{sec:CE} introduces the two main channel estimation techniques under study and the figures of merit used in this paper. Section\,\ref{sec:Results} contains results and discussions, providing an in-depth analysis and finding trade-offs. Finally, conclusions are drawn in Section\,\ref{sec:Conclusion}.

\section{System Model}
\label{sec:SystemModel}
\begin{figure}[t]
     \centering
     \begin{subfigure}[t]{0.37\columnwidth}
         \centering
         \includegraphics[width=\textwidth]{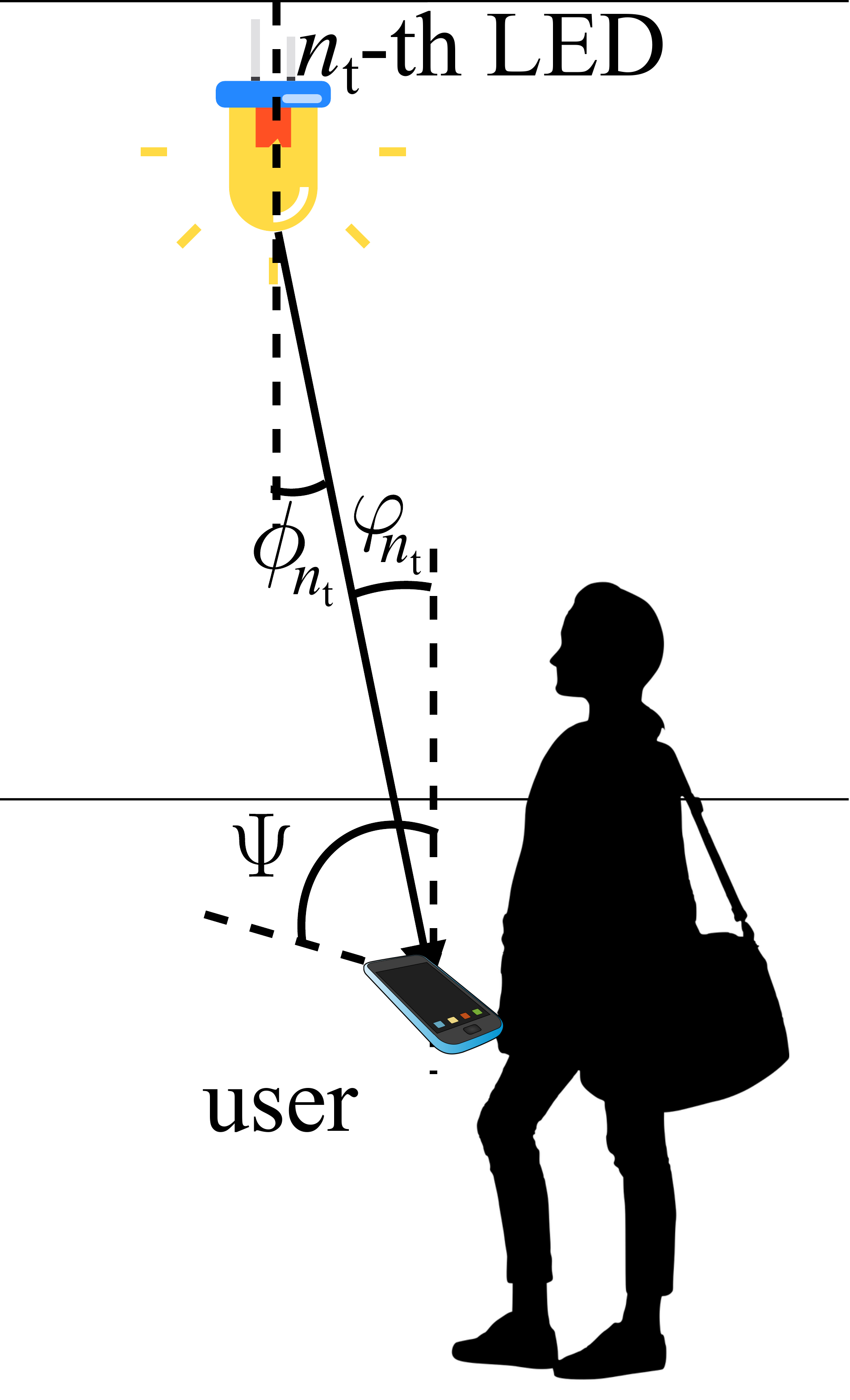}
         \caption{LoS}
         \label{fig:ScenarioLoS}
     \end{subfigure}
     \hfill
     \begin{subfigure}[t]{0.57\columnwidth}
         \centering
         \includegraphics[width=\textwidth]{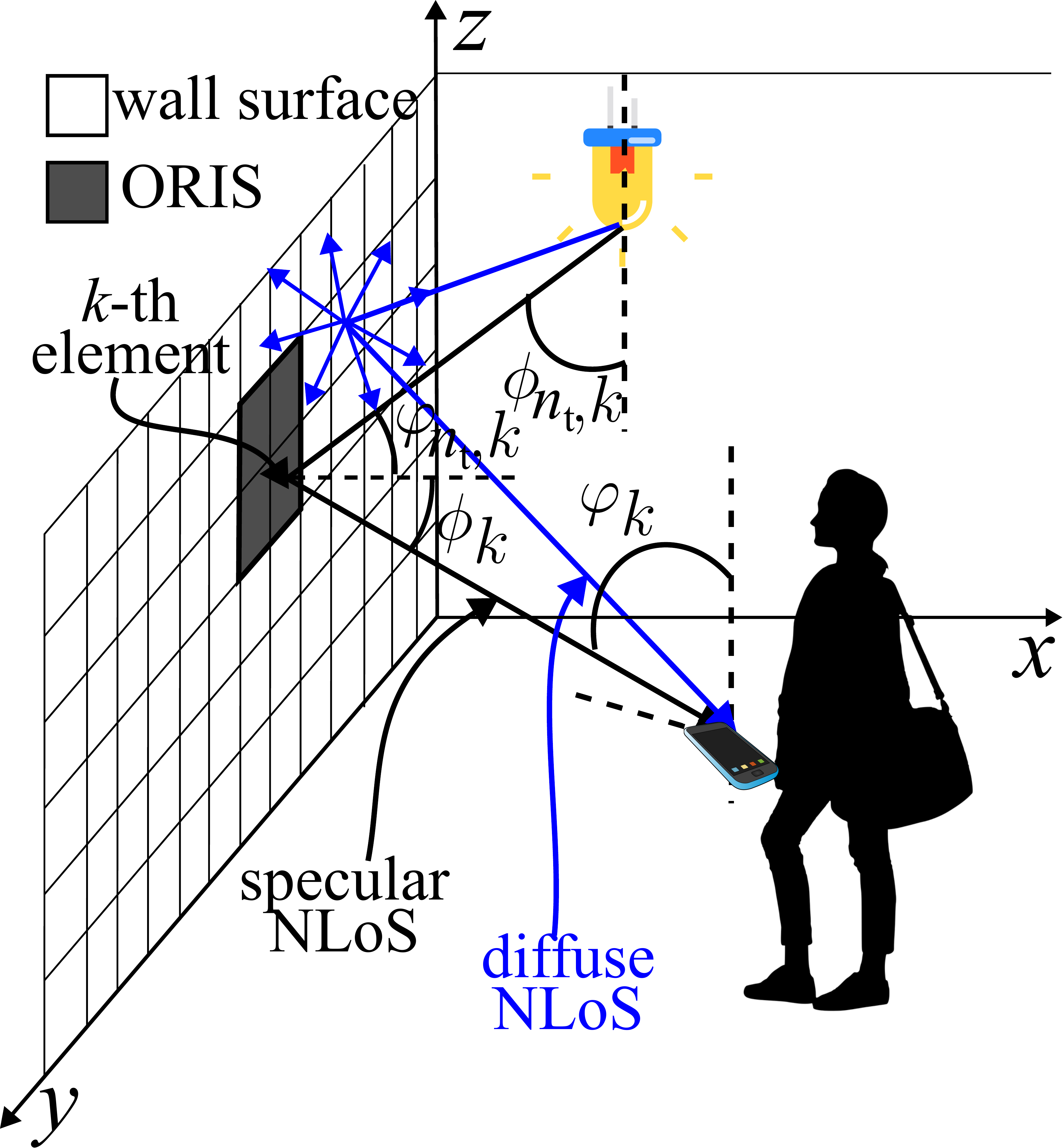}
         \caption{NLoS}
         \label{fig:ScenarioNLoS}
     \end{subfigure}
        \caption{3D illustration of LoS and NLoS propagation.}        
        \label{fig:model}
        \vspace{-6mm}
\end{figure}

We consider an indoor VLC system configuration equipped with $N_{\rm t}$ distributed optical transmitters, indexed by $n_{\rm t} \in \{0, \dots, N_{\rm t}-1\}$. The system utilizes a joint transmission scheme, where all LEDs synchronously transmit identical data streams to the target receiver. We model the LEDs as point sources emitting a Lambertian pattern, and the receiver as a point detector. The analysis focuses on a single user located at a random position, modeled by a uniform distribution within the room. This system is easily scalable to multi-user scenarios through orthogonal scheduling methods like time division multiple access (TDMA). We assume a DCO-OFDM transmission technique with $N$ subcarriers. The geometric setup assumes downward-facing transmitters and an upward-facing receiver. The channel model accounts for both LoS, and specular and diffuse NLoS propagation components that are detailed as follows and depicted in Fig.\,\ref{fig:model}.

\subsection{LoS channel gain}

The gain for the LoS link between the $n_{\rm t}$-th LED and the receiver is characterized by a Lambertian radiation profile \cite{VLCLoSChannel}, expressed as
\begin{equation}
\label{eq:ChannelModelLoS}
H_{n_{\rm t}}^{{\rm{LoS}}} =
\begin {cases}
\frac{{\left( {m + 1} \right) \cdot {A_{{\rm{PD}}}}}}{{2\pi d_{n_{\rm t}}^2}}{{\cos }^m}\left( {{\phi _{n_{\rm t}}}} \right)\cos \left( {{\varphi _{n_{\rm t}}}} \right) & 0 \le {\varphi _{n_{\rm t}}} \le {\Psi}\\ 
0 & \rm{otherwise}.
\end{cases}
\end{equation}
%where $m=-1/\log_2\left(\cos\left(\phi_{1/2}\right)\right)$ is the Lambertian index of the LED that models the radiation pattern defined by its half-power semi-angle $\phi_{1/2}$. The parameter $A_{{\rm PD}}$ stands for the active photodetector (PD) area, $d_{n_{\rm t}}$ is the Euclidean distance between LED $n_{\rm t}$ and the user, and $\phi_{n_{\rm t}}$ and $\varphi_{n_{\rm t}}$ are the irradiance and incidence angles, respectively, as represented in Fig.\,\ref{fig:ScenarioLoS}. The FoV semi-angle of the PD is denoted by $\Psi$.
In this model, $m$ represents the Lambertian order, calculated as $m=-1/\log_2\left(\cos\left(\phi_{1/2}\right)\right)$ based on the LED's semi-angle at half-power, $\phi_{1/2}$. The physical area of the photodetector (PD) is denoted by $A_{{\rm PD}}$, while $d_{n_{\rm t}}$ indicates the straight-line distance from the source to the user. The geometric relationship, including the irradiance angle $\phi_{n_{\rm t}}$ and the angle of incidence $\varphi_{n_{\rm t}}$, is illustrated in Fig.\,\ref{fig:ScenarioLoS}. Finally, $\Psi$ defines the PD's field-of-view (FoV) semi-angle.

\subsection{Specular NLoS channel gain}
When considering specular reflections, the link is modeled by treating the user as if they were at a virtual ``image point''. Here, the effective path length is the total distance of the LED$\rightarrow$mirror$\rightarrow$user trajectory, as shown in Fig.\,\ref{fig:ScenarioNLoS}. Following \cite{MirrorsVSMEtasurfaces}, this NLoS component is formulated as
%A specular reflection is equivalent to considering a user positioned at the image point, where the total distance is the sum of the two distances in the path LED$\rightarrow$mirror$\rightarrow$user, as represented in Fig.\,\ref{fig:ScenarioNLoS}. In this case the channel model can be formulated as~\cite{MirrorsVSMEtasurfaces}
\begin{equation}
\label{eq:ChannelModelNLoSRIS}
H_{n_{\rm t},k}^{{\rm{ORIS}}} =
\begin {cases}
\hspace{-1mm}{{\hat{r}} {\cdot} {\frac{{\left( {m + 1} \right) \cdot {A_{{\rm{PD}}}}}}{{2\pi {{\left( {{d_{n_{\rm t},k}} + {d_{k}}} \right)}^2}}}{{\cos }^m}{\left( {{\phi _{n_{\rm t},k}}} \right)}{\cos} \left( {{\varphi _{k}}} \right)}} & { 0 {\le} {\varphi _{k}} {\le} {\Psi}} \\
\hspace{-1mm}0 & \rm{otherwise}, \\
\end{cases}
\end{equation}
where $\hat{r}$ denotes the reflectivity of the specular surface (such as an ORIS). The variables $d_{n_{\rm t},k}$ and $d_{k}$ represent the distances from the LED to the $k$-th reflecting element and from that element to the receiver, respectively. By utilizing a MEMS-based structure, the ORIS can adjust its orientation to steer the reflected beam, ensuring the maximum incident power is directed toward the user’s location. The irradiance angle toward the reflector is $\phi _{n_{\rm t},k}$, and the angle of arrival at the user from the reflector is $\varphi_{k}$.

%where $\hat{r}$ stands for the reflection coefficient of a specular surface (e.g. ORIS), $d_{n_{\rm t},k}$ and $d_{k}$ are the Euclidean distance from LED $n_{\rm t}$ to reflector element $k$, and from reflector element $k$ to the user, respectively. Due to its installation into a MEMS structure, the ORIS orientation mobility controls the reflected path direction so that the impinging light power is forwarded to the direction where the user is located. The irradiance angle from LED $n_{\rm t}$ to reflector element $k$ is represented by $\phi _{n_{\rm t},k}$, and the incident angle from ORIS element $k$ to user is represented by $\varphi _{k}$.

\subsection{Diffuse NLoS channel gain}
As depicted in Fig.\,\ref{fig:ScenarioNLoS}, light hitting a standard wall surface scatters in numerous directions. This diffuse channel gain is typically modeled as \cite{VLCLoSModel}
%As Fig.\,\ref{fig:ScenarioNLoS} shows, the light impinging onto a wall surface is reflected in multiple directions. The channel gain can be modeled as~\cite{VLCLoSModel}
\begin{equation}
\label{eq:ChannelModelNLoSdiff}
\hspace{-5mm}
H_{n_{\rm t},k}^{{\rm{wall}}} {=}\hspace{-1mm}
\begin {cases}
\hspace{-1mm} {{\tilde{r}}  \frac{{\left( {m + 1} \right)  {A_{{\rm{PD}}}}}}{{2{\pi}d_{n_{\rm t},k}^2d_{k}^2}}\hspace{-1mm}{A_k}{{\cos }^m}\hspace{-0.5mm}{\left( {{\phi _{n_{\rm t},k}}} \right)}{\cos \hspace{-0.5mm}\left( {{\varphi _{n_{\rm t},k}}} \right)}} { {\cos \hspace{-0.5mm}{\left( {{\phi _{k}}} \right)}}{\cos \hspace{-0.5mm}{\left( {{\varphi _{k}}} \right)}}} \\& {\hspace{-3cm} 0 {\le} {\varphi _{k}} {\le} {\Psi}} \\
\hspace{-1mm} 0 & \hspace{-3cm}\rm{otherwise}, \\
\end{cases}
\end{equation}
where $\tilde{r}$ and $A_k$ stand for the reflection coefficient and the surface area of the $k$-th wall element, respectively. Here, $\varphi_{n_{\rm t},k}$ identifies the angle of incidence at the wall, while $\phi_{k}$ is the irradiance angle from the wall toward the user. Notably, unlike the specular case, the denominators in \eqref{eq:ChannelModelNLoSRIS} and \eqref{eq:ChannelModelNLoSdiff} show that for diffuse reflections, the individual path segments are multiplicative rather than additive.

%where $\tilde{r}$ is the reflection coefficient of the wall surface element $k$ with area $A_k$. The angles $\varphi_{n_{\rm t},k}$ and $\phi_{k}$ stand for the incident and irradiance angles in the surface element $k$, respectively. In contrast to the specular reflection, it is worth noticing that the distances are multiplied instead of added (see denominators in \eqref{eq:ChannelModelNLoSRIS} and \eqref{eq:ChannelModelNLoSdiff}).

\subsection{Overall channel gain}
The overall VLC channel gain of a user can be formulated as
\begin{IEEEeqnarray}{rCl}
{H_{\rm tot}} {=} \sum\limits_{n_{\rm t}}\left( H_{n_{\rm t}}^{{\rm{LoS}}} + \sum\limits_{k\in\mathcal{K}^{\rm ORIS}} \hspace{-3mm} b_k {H_{n_{\rm t},k}^{{\rm{ORIS}}}} + \sum\limits_{k\in\mathcal{K}^{\rm wall}} \hspace{-2mm} {H_{n_{\rm t},k}^{{\rm{wall}}}}\right),
\label{eq:ChannelModel}
\end{IEEEeqnarray}
where $\mathcal{K}^{\rm ORIS}$ and $\mathcal{K}^{\rm wall}$ are the set of ORIS and wall elements, respectively. We assume that every ORIS element is composed of a number of co-located ORIS sub-elements that corresponds to the same number of optical transmitters ($N_{\rm t}$), where each ORIS sub-element is associated to a different transmitter. The random variable $b_k$ follows a Bernoulli distribution that models the orientation accuracy of the $k$-th ORIS element. Specifically, $b_k = 0$ represents an orientation failure of the ORIS element with probability $p$, while $b_k = 1$ indicates that the ORIS element is successfully aligned with the target direction with probability $1-p$ .

\section{Channel estimation in VLC}
\label{sec:CE}
Considering the overall channel gain introduced in Section\,\ref{sec:SystemModel}, and assuming an ideal channel estimation, i.e., all power is dedicated to data symbols and the channel is assumed to be known, the ideal SNR can be formulated as
\begin{IEEEeqnarray}{rCl}
\overline{\gamma} & = & \frac{{{{\left( {{{\rho} P_{\rm opt} H_{\rm tot}} } \right)}^2}}}{{{N_0}B}},
\label{eq:IdealSNR}
\end{IEEEeqnarray}
where the term $\rho$ represents the PD responsivity, while $P_{\rm opt}$ denotes the level of optical power emitted by every LED. Furthermore, $N_0$ characterizes the power spectral density of the additive white Gaussian noise (AWGN) encountered at the receiver, and $B$ is the communication bandwidth. In practice, part of the transmit power must be dedicated to pilot symbols, which decreases the actual SNR. Two main channel estimation techniques have been considered in the literature, namely PSAM and ST, whose characteristics were introduced in Section\,\ref{sec:Intro}. The actual SNR received when using each of these channel estimation methods are~\cite{ST_VLC1}
\vspace{-2mm}
\begin{IEEEeqnarray}{rCl}
\overline{\gamma}_{\rm PSAM} & = & \frac{(N-2)\cdot (1-\alpha)\cdot M}{\frac{1}{\overline{\gamma}}(L\cdot N_{\rm t} + M\cdot (N-2)\cdot(1-\alpha))},
\label{eq:SNR_PSAM}
\end{IEEEeqnarray}
and
\vspace{-2mm}
\begin{IEEEeqnarray}{rCl}
\overline{\gamma}_{\rm ST} & = & \frac{\alpha \cdot (N-2)\cdot (1-\alpha)\cdot M}{L\cdot N_{\rm t}\cdot \left( \alpha + \frac{1}{\overline{\gamma}} \right) + (1-\alpha)\cdot M \cdot (N-2)\cdot \frac{1}{\overline{\gamma}}},
\label{eq:SNR_ST}
\end{IEEEeqnarray}
where $N$ is the number of subcarriers, $\alpha$ is the ratio between the power reserved for data symbols and the total power, also denoted as the data power allocation factor and whose range is $0 < \alpha < 1$. The number of OFDM symbols during which the channel does not change is represented with $M$, and $L$ accounts for the number of samples in the maximum channel delay spread. When the channel is quasi-static, we have a large coherence time that leads to a large $M$ value, and then $\overline{\gamma}_{\rm PSAM}$ and $\overline{\gamma}_{\rm ST}$ asymptotically behave as
\vspace{-3mm}
\begin{IEEEeqnarray}{rCl}
 \lim_{M \to \infty}\overline{\gamma}_{\rm PSAM} & = & \overline{\gamma},
\label{eq:SNR_PSAMLargeM}
\end{IEEEeqnarray}
and
\vspace{-3mm}
\begin{IEEEeqnarray}{rCl}
\lim_{M \to \infty}\overline{\gamma}_{\rm ST} & = & \alpha\overline{\gamma},
\label{eq:SNR_STLargeM}
\end{IEEEeqnarray}
where PSAM approximates the ideal case regardless of $\alpha$, and ST improves with $\alpha$.

Given these SNR equations, we can compute the spectral efficiency in the ideal case as,
\vspace{-2mm}
\begin{IEEEeqnarray}{rCl}
C_{\rm Ideal} & = & \frac{1}{N}\sum\limits_{n=1}^{N/2-1}\log_2(1+\frac{e}{2\pi}\overline{\gamma}).
\label{eq:SE_ideal}
\end{IEEEeqnarray}
Note that, unlike RF systems, optical intensity modulation and direct detection (IM/DD) data transmission cannot be bounded by the Shannon capacity. Then, we followed up with the tighter bound proposed in~\cite{6636053}. In the case of PSAM and ST, the lower bound for the spectral efficiency can be formulated as
\vspace{-1mm}
\begin{IEEEeqnarray}{rCl}
C_{\rm PSAM} & = & \frac{1}{N}\sum\limits_{n\in\mathcal{N_{\rm D}}}\log_2(1+\frac{e}{2\pi}\overline{\gamma}_{\rm PSAM}),
\label{eq:SE_PSAM}
\end{IEEEeqnarray}
where $\mathcal{N_{\rm D}}$ corresponds to the set of subcarriers carrying data, and
\vspace{-2mm}
\begin{IEEEeqnarray}{rCl}
C_{\rm ST} & = & \frac{1}{N}\sum\limits_{n=1}^{N/2-1}\log_2(1+\frac{e}{2\pi}\overline{\gamma}_{\rm ST}),
\label{eq:SE_ST}
\end{IEEEeqnarray}
respectively.

\setlength{\tabcolsep}{4pt}
\begin{table}[!t]
\vspace{0.05in}
\caption{Simulation parameters.}
\label{tab:SystemParameters}
\centering
{\footnotesize
\begin{tabular}{llcr}
\hline
\hline
Notation & Parameter description & Value & Unit\\
\hline
%$d_{{\rm v}}$ & \makecell[l]{Vertical distance between \\ AP and PD} & 2.25 & [m]\\
%- & Room dimensions &  6 x 6 x 3 & [m]\\
%- & x coordinates of APs &  \makecell[c]{0.75; 2.25;\\ 3.75; 5.25} & [m]\\
%- & y coordinates of APs &  \makecell[c]{0.75; 2.25;\\ 3.75; 5.25} & [m]\\
- & LED height &  3 & [m]\\
- & User device height &  1 & [m]\\
%$P_{{\rm opt,tx}}$  & Optical transmit power per AP &  7 & [W]\\
$P_{\rm opt}$ & Optical transmit power per LED &  5 & [W]\\
$\phi_{1/2}$ & \makecell[l]{Half-power semi-angle of \\ the LED} &  70 & [deg.]\\
$\tilde{r}$ & Reflection wall coefficient & 0.4 & [-]\\
$\hat{r}$ & Reflection ORIS coefficient & 0.95 & [-]\\
%$\rho _{{\rm{ORIS}}}$ & Reflection coefficient of ORIS & 0.99 & [-]\\
%$T\left( \varphi_{k,u}  \right)$ & Optical filter gain &  1 & [-]\\
%$G\left( \varphi_{k,u}  \right)$ & Concentrator gain &  1 & [-]\\
$\rho$ & PD responsivity & 1 & [A/W]\\
$A_{\mathrm{PD}}$ & PD physical area & 1 & [cm$^2$]\\
$\Psi$ & FoV semi-angle of the PD & 40 & [deg.]\\
$B$ & Communication bandwidth & 5 & [MHz]\\
$N_0$ & \makecell[l]{Power spectral density of\\ the AWGN} & $6.25{\cdot}10^{-21}$ & [W/Hz]\\
$L$ & \makecell[l]{Number of samples of maximum\\ delay spread} & 16 & [-]\\
$N$ & Number of subcarriers & 256 & [-]\\
$M$ &  \makecell[l]{Number of OFDM symbols transmitted\\within the
coherence time} & [2, 300] & [-]\\
\hline
\hline
\end{tabular}}
\vspace{-0.5cm}
\end{table}
\setlength{\tabcolsep}{6pt}

\begin{figure}[t]
\centering
     \begin{subfigure}[t]{\columnwidth}
         \centering
         \includegraphics[width=\textwidth]{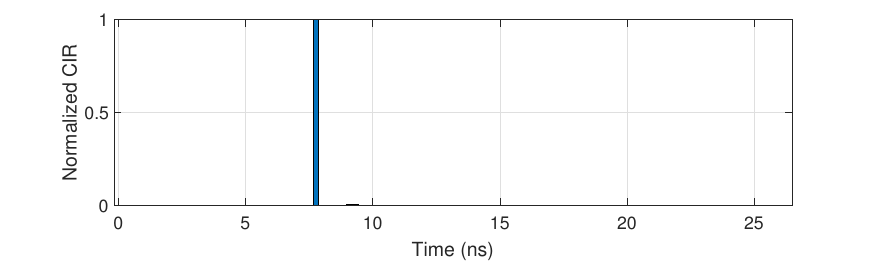}
         \vspace{-6mm}
         \caption{None ORIS}
         \label{fig:CIRnoORIS}
     \end{subfigure}
     \hfill
      \begin{subfigure}[t]{\columnwidth}
         \centering
         \includegraphics[width=\textwidth]{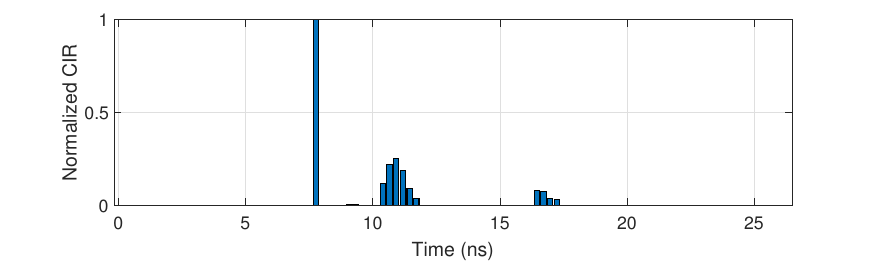}
         \vspace{-6mm}
         \caption{Small ORIS}
         \label{fig:CIRsmallORIS}
     \end{subfigure}
     \hfill
     \begin{subfigure}[t]{\columnwidth}
         \centering
         \includegraphics[width=\textwidth]{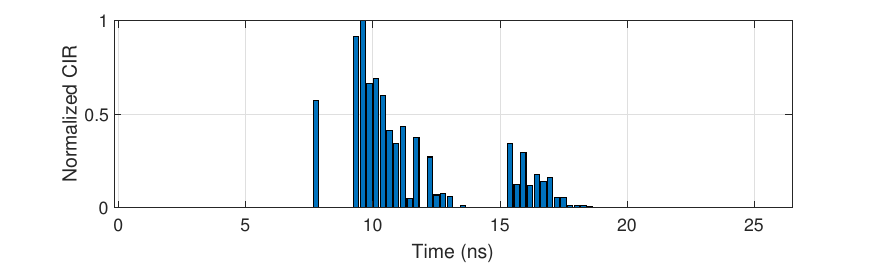}
         \vspace{-6mm}
         \caption{Medium ORIS}
         \label{fig:CIRmediumORIS}
     \end{subfigure}
     \hfill
     \begin{subfigure}[t]{\columnwidth}
         \centering
         \includegraphics[width=\textwidth]{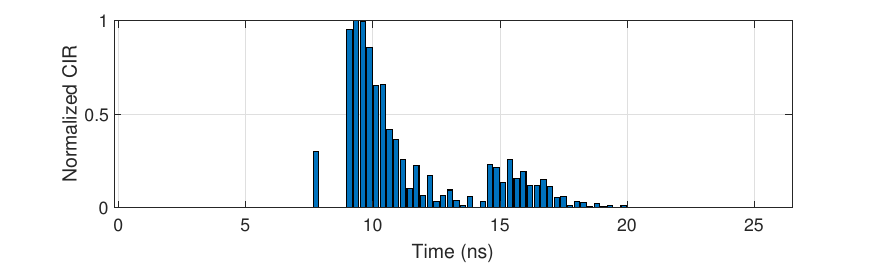}
         \vspace{-6mm}
         \caption{Large ORIS}
         \label{fig:CIRlargeORIS}
     \end{subfigure}
    \caption{Normalized CIR obtained in coordinates [0.5, 2, 1] when installing different ORIS array sizes.}
    \label{fig:CIR}
    \vspace{-7mm}
\end{figure}

\vspace{-2mm}
\section{Results}
\label{sec:Results}
%In this section, we present comprehensive simulation results for the proposed ORIS-aided VLC system with channel estimation. Without loss of generality, we consider a residential room of 4\,x\,4\,x\,3\,m, with a total of $N_{\rm t}=4$ LEDs deployed in a symmetric 2-by-2 lattice with coordinates [1, 1], [1, 3], [3, 1] and [3, 3]\,m. Though we consider NLoS contributions from the four walls, only the wall nearest to the user is the one contributing with ORIS capabilities. Each wall is divided into $K=30\times15=450$ elements. 

\begin{figure}[t]
\centering
\includegraphics[width=0.75\columnwidth]{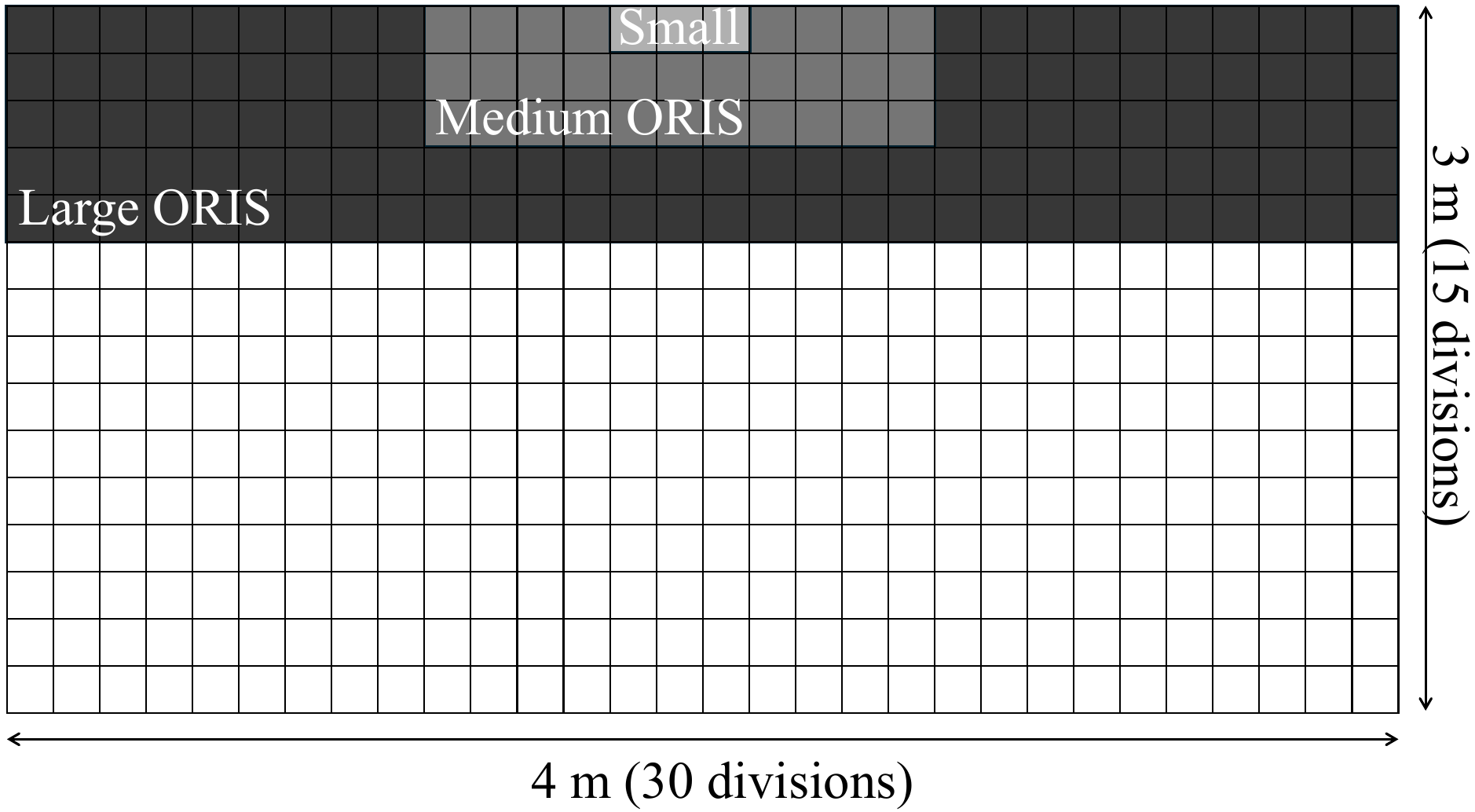}
\vspace{-2mm}
        \caption{Different ORIS array sizes formed by multiple ORIS elements considered in a wall.}
\label{fig:ORISsizes}
\vspace{-6mm}
\end{figure}

\begin{figure}[t]
\centering
\includegraphics[width=0.9\columnwidth]{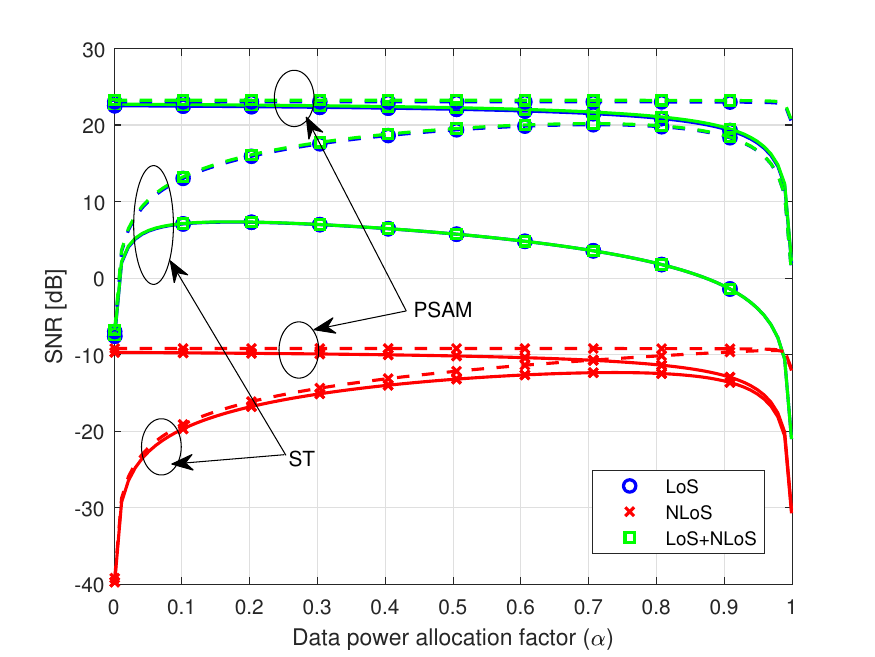}
\vspace{-2mm}
\caption{SNR obtained in coordinates [0.5, 2, 1] when PSAM or ST are implemented without ORIS elements, versus data power allocation factor $\alpha$. Solid line: $M=2$ OFDM symbols; Dashed line: $M=300$ OFDM symbols.}
\label{fig:SNRSTPSAM1user}
\vspace{-0.7cm}
\end{figure}

\begin{figure}[t]
     \centering
     \begin{subfigure}[t]{0.49\columnwidth}
         \centering
         \includegraphics[width=\textwidth]{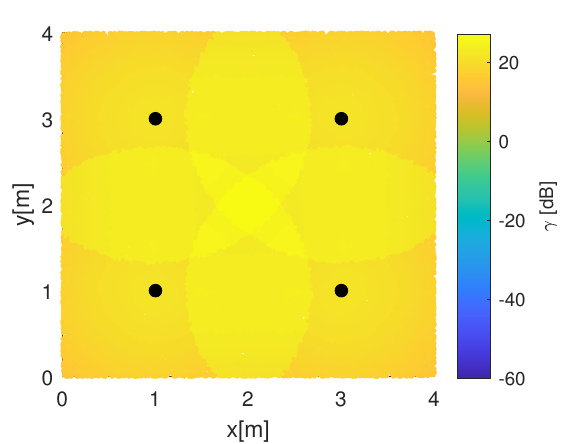}
         \caption{LoS}
         \label{fig:SNRDistLoS}
     \end{subfigure}
     \hfill
     \begin{subfigure}[t]{0.49\columnwidth}
         \centering
         \includegraphics[width=\textwidth]{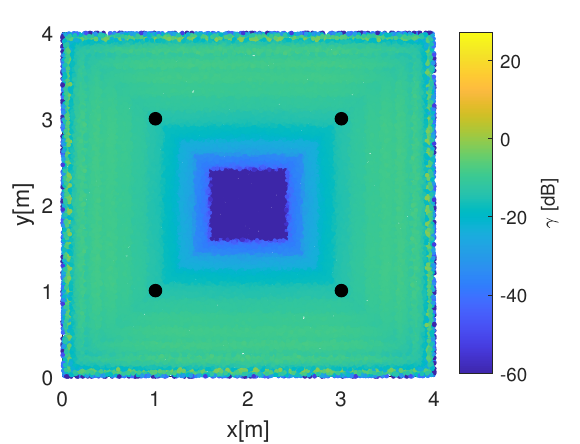}
         \caption{NLoS}
         \label{fig:SNRDistNLoS}
     \end{subfigure}
     \vspace{-1mm}
        \caption{Distribution of the ideal SNR without ORIS elements.}        
        \label{fig:SNRdist}
    \vspace{-0.5cm}
\end{figure}

This section provides a thorough performance analysis of the introduced ORIS-assisted VLC architecture, accounting for the effects of channel estimation. For our simulation setup, we evaluate a standard 4\,x\,4\,x\,3\,m indoor environment. Within this space, $N_{\rm t}=4$ LEDs are positioned in a balanced 2-by-2 grid at the following coordinates: [1, 1], [1, 3], [3, 1], and [3, 3]\,m. While the model incorporates NLoS reflections from all four walls, ORIS functionality is specifically assigned to the wall closest to the user. Each reflective surface is discretized into $K=450$ individual segments ($30\times15$).

\begin{comment}
\begin{figure}[t]
\centering
\includegraphics[width=1\columnwidth]{figs/CDF_SE_nonblocked_v2.eps}
\vspace{-4mm}
\caption{CDF of spectral efficiency obtained in the ideal case, with PSAM and ST using different $\alpha$ values and $M=300$, assuming that there is LoS.}
\label{fig:CDFSE_nonblocked}
\vspace{-5mm}
\end{figure}

\begin{figure}[t]
\centering
\includegraphics[width=1\columnwidth]{figs/CDF_SE_blocked_v2.eps}
\vspace{-4mm}
\caption{CDF of spectral efficiency obtained in the ideal case, with PSAM and ST using different $\alpha$ values and $M=300$, assuming that the user is blocked, i.e., there is NLoS.}
\label{fig:CDFSE_blocked}
\vspace{-5mm}
\end{figure}
\end{comment}

\begin{figure}[t]
\centering
     \begin{subfigure}[t]{\columnwidth}
         \centering
         \includegraphics[width=\textwidth]{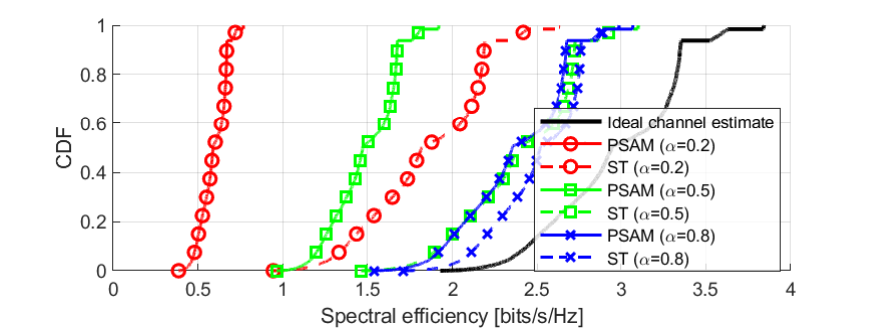}
         \vspace{-6mm}
         \caption{LoS}
         \label{fig:CDFSE_nonblocked}
     \end{subfigure}
     \hfill
      \begin{subfigure}[t]{\columnwidth}
         \centering
         \includegraphics[width=\textwidth]{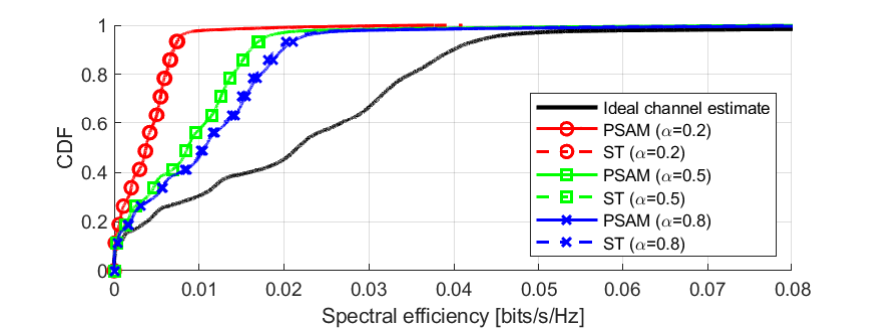}
         \vspace{-6mm}
         \caption{NLoS}
         \label{fig:CDFSE_blocked}
     \end{subfigure}
    \caption{CDF of spectral efficiency obtained in the ideal case without ORIS elements, with PSAM and ST using different $\alpha$ values and $M=300$, assuming that the user is in LoS or NLoS.}
    \label{fig:CDFSE}
    \vspace{-7mm}
\end{figure}

The specific configurations used for these simulations are detailed in Table\,\ref{tab:SystemParameters}. We have calibrated the half-power semi-angle, $\phi_{1/2}$, primarily to meet lighting standards, ensuring the beam is sufficiently broad to maintain uniform illumination throughout the room. 
%Detailed simulation parameters are shown in Table\,\ref{tab:SystemParameters}. We select the $\phi_{1/2}$ value to preferably satisfy illumination requirements, i.e., wide enough to support the lighting uniformity. 
Note that the communication bandwidth and PD physical area selected for our simulations are 5\,MHz and 1\,cm$^2$, respectively, which are within the ranges of actual systems~\cite{DLPerformanceAttocell} and aligned with standard 3-dB bandwidths of LEDs. Besides, note that the simulation parameters allow us to work in a small detector regime that, together with our LED point-source assumption and the neglected inter-element blockage, allows us to apply the specular reflection channel model formulated in \eqref{eq:ChannelModelNLoSRIS}.

%\begin{figure}[t]
%\centering
%\includegraphics[width=0.95\columnwidth]{figs/SNR_PSAM1user.eps}
%        \caption{SNR obtained in coordinates [0.5, 2, 1] when PSAM is implemented, versus data power allocation factor $\alpha$. Solid line: $M=2$ OFDM symbols; Dashed line: $M=300$ OFDM symbols.}
%\label{fig:SNRPSAM1user}
%\vspace{-0.6cm}
%\end{figure}

%\begin{figure}[t]
%\centering
%\includegraphics[width=0.95\columnwidth]{figs/SNR_ST1user.eps}
%\vspace{-1mm}
%\caption{SNR obtained in coordinates [0.5, 2, 1] when ST is implemented, versus data power allocation factor $\alpha$. Solid line: $M=2$ OFDM symbols; Dashed line: $M=300$ OFDM symbols.}
%\label{fig:SNRST1user}
%\vspace{-0.6cm}
%\end{figure}

We consider a DCO-OFDM transmission system with $N$=256 subcarriers. Fig.\,\ref{fig:CIR} shows the channel impulse response (CIR) of a user located at position [0.5, 2, 1] with 4 different cases of ORIS deployed: no ORIS, small ORIS array size (0.08\,m$^2$), medium ORIS array size (0.88\,m$^2$), and large ORIS array size (4\,m$^2$), as depicted in Fig.\,\ref{fig:ORISsizes}. In this case we consider that all ORIS elements successfully point towards the target user. As can be seen, the larger the ORIS size the more contributions coming from reflections. Note that when the ORIS array size is large enough, contributions from reflections overcome contribution coming from the LoS path, which corresponds to the signal that arrives first. We can see that the maximum delay spread is around 26\,ns. Considering a sampling rate of 650\,MHz, we have a resulting $L=16$ samples of the maximum delay spread in the system, which is also shown in Table\,\ref{tab:SystemParameters}.

\subsection{Results obtained without ORIS elements}
Fig.\,\ref{fig:SNRSTPSAM1user} shows the SNR received by a user located at the coordinates [0.5, 2, 1], when using a PSAM or ST estimation method, computed as \eqref{eq:SNR_PSAM} and \eqref{eq:SNR_ST}, respectively. In the case of PSAM, we can see an SNR decrease of around 30\,dB when there is NLoS. Besides, when there is LoS, the contribution coming from reflections (NLoS) is negligible. We also observe that the number of OFDM symbols to average is not so critical in PSAM, as similar results are obtained in the case of $M=2$ and $M=300$, except for large $\alpha$ values, where the power dedicated to pilots is small and the noise is comparatively large. In this case, averaging a large number of OFDM symbols offers a constant SNR that is very close to the ideal SNR $\overline{\gamma}$, as shown  in \eqref{eq:SNR_PSAMLargeM}.

\begin{figure*}
     \centering
     \begin{subfigure}[t]{0.325\textwidth}
         \centering
         \includegraphics[width=\textwidth]{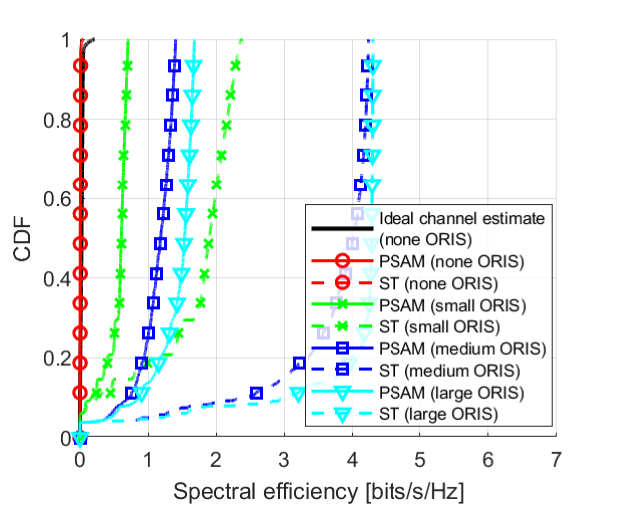}
         \vspace{-6mm}
         \caption{$\alpha=0.2,\,M=300$}
         \label{fig:alpha02M300}
         \vspace{-0.5mm}
     \end{subfigure}
     \hfill
     \begin{subfigure}[t]{0.325\textwidth}
         \centering
         \includegraphics[width=\textwidth]{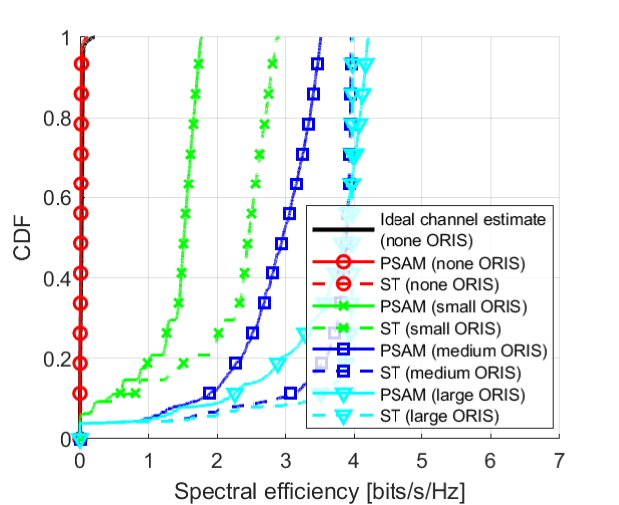}
         \vspace{-6mm}
         \caption{$\alpha=0.5,\,M=300$}
         \label{fig:alpha05M300}
         \vspace{-0.5mm}
     \end{subfigure}
     \hfill
     \begin{subfigure}[t]{0.325\textwidth}
         \centering
         \includegraphics[width=\textwidth]{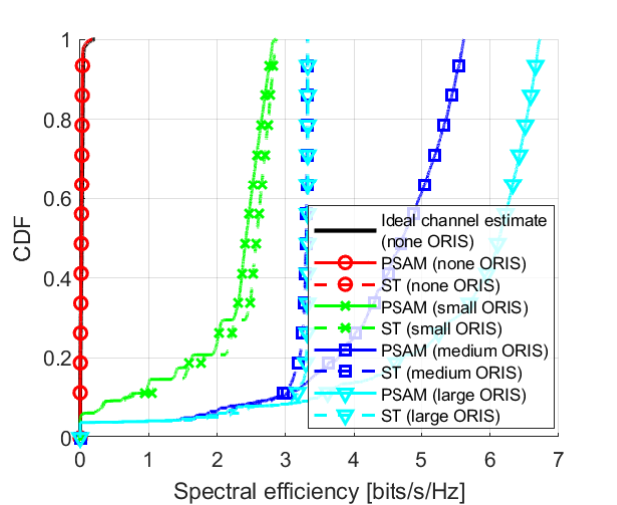}
         \vspace{-6mm}
         \caption{$\alpha=0.8,\,M=300$}
         \label{fig:alpha08M300}
         \vspace{-0.5mm}
     \end{subfigure}
     \newline
          \begin{subfigure}[t]{0.325\textwidth}
         \centering
         \includegraphics[width=\textwidth]{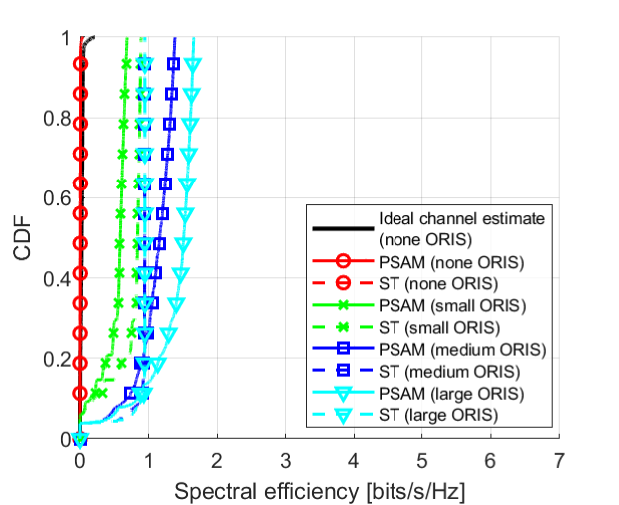}
         \vspace{-6mm}
         \caption{$\alpha=0.2,\,M=2$}
         \label{fig:alpha02M2}
     \end{subfigure}
     \hfill
     \begin{subfigure}[t]{0.325\textwidth}
         \centering
         \includegraphics[width=\textwidth]{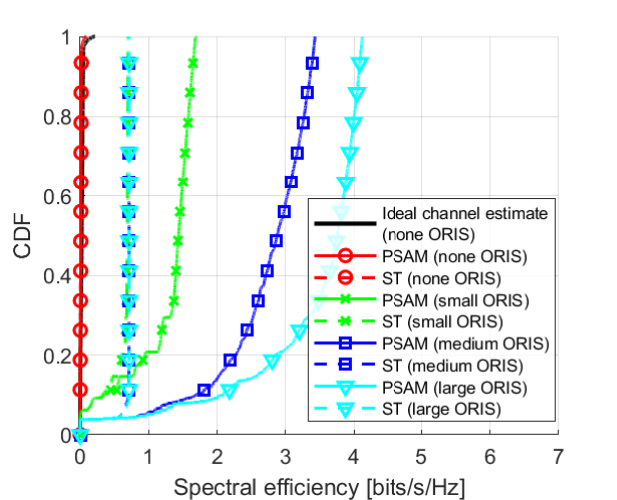}
         \vspace{-6mm}
         \caption{$\alpha=0.5,\,M=2$}
         \label{fig:alpha05M2}
     \end{subfigure}
     \hfill
     \begin{subfigure}[t]{0.325\textwidth}
         \centering
         \includegraphics[width=\textwidth]{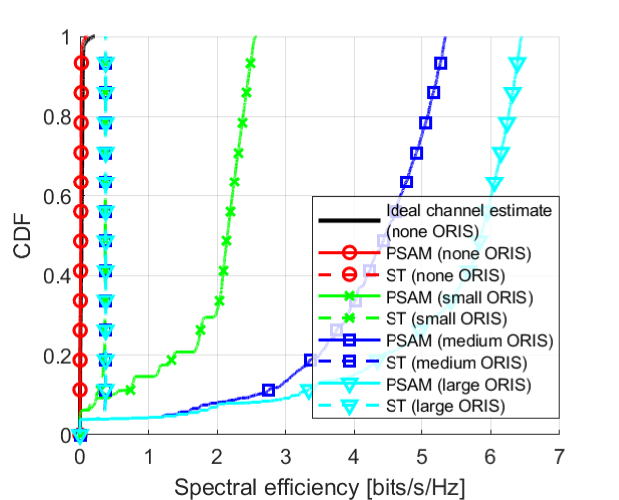}
         \vspace{-6mm}
         \caption{$\alpha=0.8,\,M=2$}
         \label{fig:alpha08M2}
     \end{subfigure}
     \vspace{-2mm}
        \caption{Spectral efficiency for different $\alpha$ and $M$ values when PSAM and ST channel estimation techniques are used, evaluated under different sizes of ORIS array and compared to no-ORIS case.}
         \label{fig:SE_AlphasandMs}
         \vspace{-5mm}
\end{figure*}

In the case of ST, we can also see in Fig.\,\ref{fig:SNRSTPSAM1user} an SNR decrease of around 30\,dB when there is NLoS and $M$ is large enough (dashed lines). In this case, we see that SNR improves with $\alpha$, which was mathematically proven in \eqref{eq:SNR_STLargeM}. However, for low $M$ values (solid lines), the SNR, in case of LoS, decreases with $\alpha$ because it is better to allocate power to the pilots for estimation as the SNR is large enough for correctly receiving data. Differently, in case of NLoS and low $M$ values, it does not matter how many $M$ OFDM symbols are averaged, as the channel is very bad, so it is preferable to allocate more power to data by configuring a large $\alpha$.

Let us now analyze the SNR obtained at any room location. Fig.\,\ref{fig:SNRdist} shows the ideal SNR $\overline{\gamma}$ in case of LoS and NLoS. We can see as the SNR obtained in case of LoS is around 20\,dB at any location. However, if the user suffers a LoS blockage and exclusively depends on NLoS links, the received SNR is around -10\,dB at positions close to walls, whereas it is even lower at positions centered in the room. These graphs clearly show the damage that LoS blockage creates in SNR, which means a decrease of around 30\,dB in SNR that can be dramatic for communication and may lead to outages.

We then analyze this effect in communication by evaluating the spectral efficiency as formulated in \eqref{eq:SE_ideal}, \eqref{eq:SE_PSAM} and \eqref{eq:SE_ST} for the case of ideal SNR, SNR with PSAM, and SNR with ST, respectively. Fig.\,\ref{fig:CDFSE_nonblocked} shows the cumulative distribution function (CDF) of the spectral efficiency in the case of LoS, when a single user is located at a random position uniformly distributed within the room, for different $\alpha$ values and $M=300$. Generally, the larger the $\alpha$ the better the spectral efficiency is as it was also shown in Fig.\,\ref{fig:SNRSTPSAM1user}. At low $\alpha$ values ($\alpha=0.2$ and $\alpha=0.5$), ST offers better results than PSAM as it uses all subcarriers for both data and pilot transmissions, while PSAM sacrifices entire subcarriers for pilot transmission, which impacts the spectral efficiency. Differently, at large $\alpha$ values ($\alpha=0.8$) PSAM dedicates only a few subcarriers for pilot transmission, which is enough for a good channel estimation due to having a large SNR (LoS existence) and does not compromise the spectral efficiency. In this case, ST is affected by the low power dedicated to pilots in all subcarriers. Fig.\,\ref{fig:CDFSE_blocked} shows the CDF of the spectral efficiency in the case of NLoS for different $\alpha$ values and $M=300$. In this case, as the SNR is very low due to the LoS blockage, both ST and PSAM estimation techniques provide a similar spectral efficiency. Still, the larger the $\alpha$ the better the spectral efficiency as the channel estimation will be equally bad, regardless of the power allocated to pilot transmission. Then, it is better to allocate more power to data symbols.

\subsection{Results obtained in the proposed ORIS-aided scenario}

As seen in Fig.\,\ref{fig:CDFSE_blocked}, and when we compare it with respect to results of Fig.\,\ref{fig:CDFSE_nonblocked}, suffering LoS blockage dramatically impacts the spectral efficiency. Besides, the difference with respect to the results obtained with an ideal channel estimate is even larger. This clearly calls for novel techniques that allow increasing the received SNR for a proper channel estimation and, as a consequence, an improved spectral efficiency.

We then consider the deployment of ORIS arrays in the room with different sizes represented in Fig.\,\ref{fig:ORISsizes} and we evaluate their performance. The CDF of the spectral efficiency for different combinations of $\alpha$ and $M$ values under this new setup is represented in Fig.\,\ref{fig:SE_AlphasandMs}. %Namely, we compare 4 different cases of ORIS deployed: no ORIS, small ORIS array size, medium ORIS array size, and large ORIS array size, as depicted in Fig.\,\ref{fig:ORISsizes}. 
Here we consider that the ORIS elements are properly oriented to contribute to the user, i.e., $p=0$ in the modeling of $b_k$ of \eqref{eq:ChannelModel}. Note that the implementation of ORIS elements means a substantial increase in SNR that allows obtaining a spectral efficiency larger than the one obtained in the ideal estimate case without an ORIS deployed. In case $M=2$, ST does not perform well as it needs a large number of OFDM symbols to average for cancelling the contamination over pilots induced by the data symbols. In the case of $M=300$, ST performs better than PSAM only at small $\alpha$ values. However, spectral efficiency is maximized at large $\alpha$ values when using PSAM, as installing ORIS elements enhances the communication channel, then producing a larger SNR in most of the room positions. This means that most of the power can be allocated for data symbols and only a little will be dedicated to pilots for channel estimation.

\begin{figure}[t]
\centering
\includegraphics[width=0.9\columnwidth]{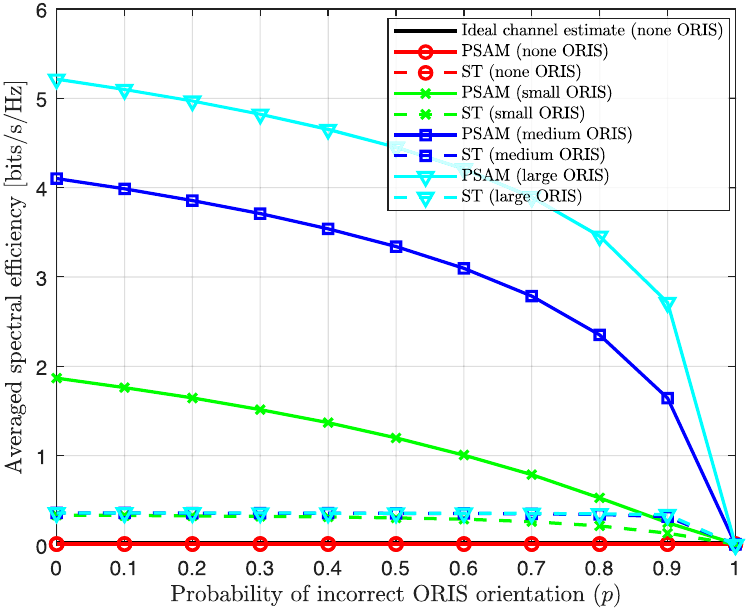}
\vspace{-1mm}
\caption{Averaged spectral efficiency for $\alpha=0.8$ and $M=2$ when PSAM and ST channel estimation techniques are used, evaluated under different probabilities of ORIS incorrectly oriented.}
\label{fig:AveragedSE_ORISmisoriented}
\vspace{-7mm}
\end{figure}

%Finally, we evaluate the spectral efficiency with an imperfect ORIS orientation, i.e., $p\neq 1$ in the modeling of $b_k$ in \eqref{eq:ChannelModel}. Fig.\,\ref{fig:AveragedSE_ORISmisoriented} shows the averaged spectral efficiency for $\alpha=0.8$ and $M=2$ when both PSAM and ST are implemented under different $p$ values. An ORIS misalignment (larger $p$ values) means a decrease in the spectral efficiency, but values of around 0.5 are still affordable as they provide a reasonable averaged spectral efficiency. This means that, with half of ORIS elements properly oriented, we could still offer a great spectral efficiency in the room.

Lastly, we assess the spectral efficiency under imperfect ORIS orientation, represented by $p > 0$ in the modeling of $b_k$ in \eqref{eq:ChannelModel}. Fig.\,\ref{fig:AveragedSE_ORISmisoriented} illustrates the averaged spectral efficiency for $\alpha=0.8$ and $M=2$ when both PSAM and ST are employed across a range of $p$ values. While ORIS misalignment inevitably leads to a reduction in spectral efficiency, the system remains robust. Even at $p \approx 0.5$, the performance is acceptable, yielding a reasonable averaged spectral efficiency. This suggests that the system can maintain high performance throughout the room even if only half of the ORIS elements are accurately oriented.

\vspace{-2mm}
\section{Conclusion}
\label{sec:Conclusion}
\vspace{-1mm}
%This paper investigates the contribution of ORIS elements in VLC systems without assuming an ideal channel estimation. We analyze the communication performance when using two main pilot-based channel estimation techniques, PSAM and ST, evaluated at different configuration parameters such as the data power allocation factor, and the averaged OFDM symbols. We show that deploying ORIS elements means a considerable SNR improvement even when LoS path is blocked. This translates into large optimal data power allocation factors in the case of PSAM, and spectral efficiency improvements with respect to the case where an ideal channel estimate without an ORIS deployment is considered. ST performs well if the number of averaged OFDM symbols is large, but it cannot overcome PSAM results when ORIS elements are implemented. Generally, spectral efficiency values of around 5\,bits/s/Hz can be easily obtained if ORIS elements are deployed and PSAM estimation technique is implemented, which is in the order of, or even larger than, the spectral efficiency values obtained in the case of LoS.

This paper evaluates ORIS-aided VLC systems under non-ideal channel estimation, comparing PSAM and ST techniques across various power allocation factors and symbol averaging. Our findings demonstrate that ORIS elements significantly boost SNR, even during LoS blockage, enabling higher optimal data power allocation and superior spectral efficiency compared to non-ORIS scenarios. Furthermore, we show that the system is robust to imperfect ORIS orientation; even with only half of the elements accurately aligned, the system maintains acceptable performance. Overall, the combination of ORIS and PSAM achieves spectral efficiency values near 5\,bits/s/Hz, performance that is comparable to, or even exceeds, standard LoS conditions.


\vspace{-1mm}
\section*{Acknowledgment}
\vspace{-1mm}

This work has been partially funded by Ramón y Cajal grant RYC2023-042518-I, funded by MCIU/AEI/10.13039/501100011033 and FSE+; by project PID2023-147305OB-C31 (SOFIA-AIR) by MCIN/ AEI/10.13039/501100011033/ERDF, UE; and by project PID2024-156038OA-I00 (HOLIFI6G) by MICIU /AEI /10.13039/501100011033 / FEDER, UE.



\vspace{-1mm}
\bibliography{./references}
\bibliographystyle{IEEEtran}

\end{document}